# Consistency Ensuring in Social Web Services Based on Commitments Structure

Marzieh Adelnia
*Department of Computer Engineering*
*University of Isfahan*
*Isfahan, Iran*
adelniya_m@yahoo.com

Mohammad Reza Khayyambashi
*Department of Computer Engineering*
*University of Isfahan*
*Isfahan, Iran*
m.r.khayyambashi@eng.ui.ac.ir

*Abstract* - Web Service is one of the most significant current discussions in information sharing technologies and one of the examples of service oriented processing. To ensure accurate execution of web services operations, it must be adaptable with policies of the social networks in which it signs up. This adaptation implements using controls called "Commitment". This paper describes commitments structure and existing research about commitments and social web services, then suggests an algorithm for consistency of commitments in social web services. As regards the commitments may be executed concurrently, a key challenge in web services execution based on commitment structure is consistency ensuring in execution time. The purpose of this research is providing an algorithm for consistency ensuring between web services operations based on commitments structure.

*Index Terms– Commitment, Social Commitment, Consistency, Social Network, Social Web Service, Web Service.*

## I. INTRODUCTION

Web is the largest transformable-information framework. Many research has been accomplished about the web and related technologies in the past two decades. Web 1.0 as a web of cognition, web 2.0 as a web of communication, web 3.0 as a web of co-operation and web 4.0 as a web of integration are introduced as four generation of the web since the advent of the web.

"*Web 2.0 was presented in 2004 as a read-write web. The technologies of web 2.0 allow assembling and handling large global crowds with common interests in social interactions*" [1].

Today, web services have become one of the most important information sharing technologies on the web and one of the examples of service oriented processing. A Web service is a software designed to support interoperable machine-to-machine interaction over a network. It has an interface described in a format that is processed by machine. Other systems communicate with the web service in a format of SOAP messages, typically conveyed using HTTP with an XML serialization in conjunction with other web-related standards. In fact a web service is an abstract notion that must be implemented by a concrete agent. The agent is the base component of software or hardware that sends and receives messages, while the service is the resource characterized by the abstract set of functionality that is provided. Although if the agent changes, the web service don't change. According to the W3C, a web service "*is a software application identified by a URI, whose interfaces and binding are capable of being defined, described, and discovered by XML artifacts and supports direct interactions with other software applications using XML based messages via Internet-based applications*".

In recent years, many users have registered in various social networks that use web services. Regarding the growth of social networks and the tendency of users for registering on them, web services can be studied from a social computing viewpoint. Social computing is an area of computer science that is concerned with the intersection of social behavior and computational systems. Social computing is basically the use of a computer for social goals. A prime example for these processes is applications based on web 2.0 like social networks and blogs.

With merging service oriented computing and social computing, social web services are produced that are more complicated than regular web services**.** To ensure accurate execution of web services operations, they must be adaptable with the policies of the social networks in which they sign up. This adaptation is implemented using controls called "Commitment". In other words, transactions between web services components and social networks lead to the creation, management and use of commitments [2].

As regards to the concurrent execution of social web services and commitments in a social network, consistency of commitments is one of the most important challenges in this topic. In previous research [3], commitments structure is analyzed and based on their attributes classified. In this paper, the algorithm is designed. In addition to this, supplemental commitments are defined to optimizing the social web services operation and the algorithm is checked with Facebook dataset.

## II. BACKGROUND

Many research exists about social web services, commitments and consistency in web services. This section provides an overview on these topics. The section is partitioned into three groups and each group is assigned to one topic.

*A. Overview on Social Web Services.*

The synergy between social computing and service oriented computing has eventuated into social web services. Existing research focuses on adoption of web services to social networks.

In 2010, Maaradji proposed a social constructor named "SoCo" that suggests and helps users for next their operations (like selecting specific web service). So users may like to perform an operation that their friends have done in social networks [4].

In the other research in 2011, Maamar et al. categorized social networks to three groups including [5]:

- Collaboration social networks. *"By emerging their respective functionalities, social Web services have the capacity to work together and response to complex user requests. In fact, a social Web service manages its own network of collaborators"*.
- Substitution social networks. *"Although social web services compete against each other, they can still help each other when they fail as long as they offer similar functionalities."*
- Competition social networks. *"Social web services compete against each other when they offer similar functionalities. Their non-functional properties differentiate them when users' non-functional requirements must be satisfied."*

Maamar also purposed an approach for weaving social networks operation using web services. The results of his research lead to creating social web services [6].

In 2013, Maamar et al. studied the social qualities that web services present at run time to identify and assign adequate social qualities to communities that host these web services. They discussed the binding of communities of web services to social qualities like selfishness. The quality of communities is presented based on how web services respond to the scenarios like collaboration, delegation, competition and coopetition [7].

*B. Overview on commitments*

Fornara and Colombetti are the first who studied and used commitments. They defined a general formula for commitments and used them for speech evaluation [8].

In 2003, Grosof et al. proposed a rule-based method for representation of e-contracting. It is named "SweetDeal" and uses XML rules and ontologies for simulating business contracting. In SweetDeal approach, a business contract is a set of activities that can be decomposed into sub-activities. The terms of contracts use a set of commitments to execute operation by agents. The algorithm uses a coordination method to manage agent activities [9].

In 2005, Bentahar et al. presented a new persuasion dialogue game for agent communication. They modeled dialogue game by a framework based on social commitments and arguments, named CAN[1]. CAN framework allows to model dynamic communication in levels of activities that agents apply to commitments and in levels of argumentation relations. This dialogue game is specified by indicating its entry, dynamic and exit conditions. They proposed a set of algorithms for the implementation of the persuasion protocol and discussed their termination, complexity and correctness [10].

In 2006, Carabelea et al. studied the one of the main challenge in multi agent system. This challenge was the coordination ensuring of autonomous agent in open heterogeneous system. They used social commitments to solve this challenge. In fact they combined two models of coordinating agent, commitment-based interactions and organizations. They described how one can use social commitments to represent the expected behavior of an agent playing a role in an organization. They defined an organizational structure as a collection of roles, where a role is considered to be the subject of different types of social commitments and policies [11].

In 2008, Narendra defined a contract as a collection of the participant's commitments toward each other. The interactions that take place in a contract are understood in terms of how they operate on the participant's commitments. The operation on a commitment cause its state changes according to a life cycle [12].

In 2009, Singh et al. advocated for examining Service-Oriented Architecture (SOA) principles from a commitment perspective. As regards *"existing service-oriented architectures are formulated in terms of low level abstractions far removed from business services"*, In CSOA[2], the components are business services and the connectors are patterns, considered as commitments, which protect key elements of service engagements. Each participant is implemented as an agent; interacting agents perform a service engagement by creating and elaborating commitments to one another [13].

In 2010, El-Menshawy et al. showed that current methods fail to capture the meaning of interactions that arise in real-life business scenarios and proofed commitments increase flexibility and intuitively in protocols. They presented an exploder definition for commitments for using in a larger level. In their definition, a new grammar named "Computation Tree Logic" (*CTL)* and terms like $SC^P$ for unconditional commitments and $SC^C$ for conditional commitments are added. *CTL* is a logical tree and commitments are the nodes of tree that are organized in the tree base on logical regulation in transaction execution time [14].

---

[1] Commitment and Argument Network
[2] Commitment - based Service Oriented Architecture

In 2012, Maamar et al. implemented social web services using commitments. They presented the architecture implementing social web services installation in terms of monitoring level, social level and service level. They introduced "Responsibility" concept in social web services and based on this concept, designed commitments for responsibility of social web services [2].

In 2013, Maamar et al. presented a set of responsibilities for social web services based on commitments and defined requirement commitments for responsibilities. A key challenge in this research is consistency ensuring between commitments in social web services that this paper has proceeded it [15].

In 2014, Sultan et al. merged knowledge and social commitments and presented a new framework to model and alter stochastic multi-agent systems. They defined a new multi model logic called $PCTL^{kc}$ [1]. The $PCTL^{kc}$ merge probabilistic logic of knowledge (PCTLK) and probabilistic logic of commitments (PCTLC) [16].

In 2015, Nardi et al. proposed a commitment-based account of the concept of service that uses a core reference ontology [17]. This mechanism called UFO-S[2]. UFO consists of three base element as follows:
1. UFO-A. It is an ontology of objects.
2. UFO-B. It is an ontology of event.
3. UFO-C. It is an ontology of social entities based on UFO-A and UFO-B.

*C. Overview on Consistency in Web Services*

In 2004, Heckel et al. discussed an approach to model consistency management for component-based architectures and its application to web service architectures. They proposed an algorithm that receives activity diagram of web service and translate it into CSP[3]. The CSP analyzes it for deadlock freedom [18].

In 2005, Greenfield et al. proposed a protocol to checking dynamic consistency. Their protocol can be run at the termination of a service-based application. The key of their work is establishing a relationship between internal service states, messages and application-level protocols. This protocol is based on the way that the reflection and transfer of critical state within messages links the local consistency expressions for each of the participating services. It should let verify global consistency at termination without needing global consistency expressions and an overall coordinator to evaluate them [19].

In 2008, Choi et al. suggested a mechanism to ensure consistency for web services transactions. This mechanism recognizes inconsistent states of transactions and replaces them with consistent states. Mechanism operation is designed by a waiting graph of web services transactions and a coordinator that check waiting graph. If coordinator is certified about deadlock lack, allows transaction to execute. Also if deadlock occurred, coordinator recognizes a safe state using waiting graph and replaces it instead deadlock state. Based on this mechanism, WTDP[4] is designed [20].

In 2011, Shan-liang discussed a model for transactions processing coordination based on BPEL. In this model a coordinator is used for web services transactions and if deadlock occurred, coordinator rollbacks operations [21].

In 2011, also Hemel et al. studied consistency checking of web applications as a problem. Regards to parts of an application are defined with separate domain-specific languages, which are not checked for consistency with the rest of the application, they presented a declarative, rule-based approach to linguistic integration and consistency checking in web application. They argued that domain-specific languages should be designed from the ground up with static verification and cross-aspect consistency checking in mind, providing linguistic integration of domain-specific sublanguages [22].

In 2014, Adelnia et al. proposed an approach for commitment classification based on commitment structure. The algorithm for consistency ensuring between commitments is designed based on attributes of commitments structure [3].

### III. COMMITMENTS DEFINITION

This section describes commitments. First commitments types are explained then the commitments structure are expanded. In this section the responsibilities are required in social networks, are presented and their commitments are defined and expounded.

*A. Commitments Types*

Two types of commitments are defined [15]:
1. Social Commitments: Responsibilities bonded by one agent to another agent. Agents are usually the web services in social networks. Social commitments guarantee the proper use of the social networks in which they sign up.
2. Business Commitments: Arise when social web services take part in compositions and guarantee the proper development of composite web services in response to users' requests.

---

[1] Probabilistic Computation Tree Logic of Knowledge and Commitments
[2] Unified Foundational Ontology - Service
[3] Communicating Sequential Processes
[4] Web service Transaction Dependency management Protocol

## B. Commitments Structure

Maamar et al. suggested a standard formula for social commitments based on Fornara's formula. Fornara and Colombetti note that *"...intuitively a social commitment is made by an agent (the debtor) to another agent (the creditor), that some fact holds or some action will be carried out (the content)"*. In addition to this formula, Maamar also presented a list of responsibilities for social web services. A commitment structure is defined as: $C_{Resp_i}$ (debtor, creditor, content [|condition]). Condition parameter is optional [2].

Social commitments defined by Maamar et al. are listed as follows [15]:

1. $Resp_1$. "Collecting any detail (d) in a social network would require indicating the purpose (p) of this collection to this detail's owner (o)". This responsibility formula can be represented as: *Permission(Collect(d, o, valid(p)))*. The commitment of this responsibility is defined as: $C_{Resp1}$ ( $sws_i$, $sws_j$, Collect( d, $sws_j$)/ valid($p_d$)).

2. $Resp_2$. "Posting any detail (d) on a social network should be correct." This responsibility formula can be represented as *Obligation( Post( d, true))*. The commitment of this responsibility is defined as: $C_{Resp2}$ ($sws_i$, $sn_{auth}$, Post( $d_{self}$ )).

3. $Resp_3$. "Collecting any detail (d) from a social network should not be tampered after information collection." This responsibility can be represented as *Obligation( not-Tamper( d, o, collection (d)))* and its commitments is defined as: $C_{Resp3}$ ($sws_i$, $sws_j$, not-Tamper( $d_{public}$, $sws_j$ ).

4. $Resp_4$. "Signing off from a social network would require the completion of all the pending assignments (ass). It can be represented as *Permission(Signoff(status(ass)))*. Sign-off is the action and status is a function that assesses the progress (e.g., ongoing, complete, and failed) of ass." And the commitment for it is defined as: $C_{Resp4}$($sws_i$, $sn_{auth}$, Sign-off()/status(ass)).

5. $Resp_5$. "Revealing any public detail (d) to the non-members (not(m)) of a social network should not be authorized indefinitely, represented as *Obligation(not-Reveal(d, o, m, collection(d)))*. not-Reveal is the action, m corresponds to the non-members of a social network, and collection is a function that checks if collecting d is approved in compliance with $Resp_1$." The commitment of this responsibility defined as: $C_{Resp5}$ ($sws_i$, $sn_{auth}$, not-Reveal($d_{public}$, o, nm)/collection($d_{public}$)).

Business commitments structure are defined like social commitments. They are expounded in [15].

## IV. SUPPLEMENTAL COMMITMENTS

As for increasing the variety of activities in social networks, commitments may be added or changed. In this section, two commitments that can optimize and complete social web services operation, are added and described as follows:

1. $Resp_6$. Sharing any information in a social network would require specifying the goal of this action. This responsibility can be defined as: *Permission(Share(Info, o, valid(p)))* and its commitment is: $C_{Resp6}$ ($sws_i$, $sws_j$, Share (Info, $sws_j$)/valid($p_{Info}$)). Permission is the function that checks the goal, *Share* is the action of social web service, *Info* is the information which would be shared in social network like a video or a text from other social networks, *o* is the writer of info like a social web service, *valid* is the method that checks purpose of social web service and *p* is the purpose of social web service. This responsibility checks the goal of social web service and if its goal is valid, allows to social web services to share information on an account.

2. $Resp_7$. Posting any activity of user in other social networks must be correct. This responsibility can be defined as: *Obligation( PostActivity( act, true))* and its commitment is: $C_{Resp7}$( $sws_i$, PostActivity( act, $sws_j$)/ valid($p_{act}$)). PostActivity is the action of social web service, *act* is the activity of the user in other social networks, $sws_i$ is the social web service that posts the activity of user in a social network and valid($p_{act}$) is the function that checks purpose of this action and if it is valid, allows the social web service to be execute.

## V. COMMITMENTS CLASSIFICATION

For achieving consistency of commitments, commitments are classified based on their effect on databases and information of social network into [3]:

1. Reader Commitments: *"this category of commitments don't change the information of database and social network and usually act as an information collector for other social networks or purpose checker in social web services. Note that the goal of social web services that use reader commitments must be valid. Also privacy must be protected."*

2. Writer Commitments: *"unlike reader commitments, this category can change the information of database and social networks. They share Information and post activity on other social networks. So writer commitments are more effective than reader commitments in social web service transactions. Like reader commitments, in writer commitments, privacy must be controlled."*

This classification of commitments is showed in Fig.1.

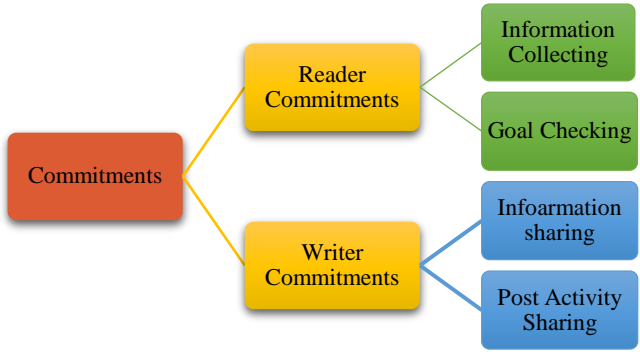

Figure 1. Commitments Classes

## VI. CONSISTENCY GUARANTOR ALGORITHM

As farther described, the action of the social web service is implemented using commitments and as regards both reader and writer commitments may be executed on social networks concurrently, a major problem that must be considered is consistency ensuring of commitments in social web services.

To designing the consistency guarantor algorithm, three concepts are considered as follows:

1. *Friend*: Commitments are *Friend* if they are reader commitments. So they are consistent in all states and database is in the safe state. It can be formulated as: $isReader(C_i) \wedge isReader(C_j) \rightarrow IsFriend(C_i , C_j)$. These commitments may collect the information of social networks or check the goal of social web services. For example if $C_{Resp1}$ and $C_{Resp3}$ are executed on an account in a social network, they are *Friend*.

   *Family*: Commitments are *Family* if they are writer commitments. In fact they effect on the state of database and information. It can be formulated as: $isWriter(C_i) \wedge isWriter(C_j) \rightarrow IsFamily(C_i , C_j)$. These commitments may share the information of social networks. For example if execute two or more $C_{Resp2}$ concurrently on an account in a social network, they are *Family*.

2. *Strange*: If commitments neither *Friend* nor *Family* are *Srange*. In this state, commitments may be reader or writer. It can be formulated as: $isReader(C_i) \wedge isWriter(C_j) \rightarrow IsStrange(C_i , C_j)$. For example if $C_{Resp1}$ and $C_{Resp2}$ are executed concurrently on an account in a social network they are *Strange*.

Since writer commitments impress database and information, so if arrival commitments of the social web services and active commitment that is running, be *Family* or *Strange*, conflict may occur. In this state, consistency must be guaranteed and if deadlock happened it would be removed and system needs to be recovered.

Suppose that a web service signs up in a social network, first authority component recognizes it. If the web service is valid in social network, it can sign up in the social network and a specific responsibility is assigned it. Also its commitments will be created based on its responsibility. This time, consistency checking between commitments is critical. Because of another commitment is executing on same account in social network, they may be inconsistent and conflict may happen. To guarantee consistency, first, condition of current commitment is checked towards active commitment that is executing. Three conditions may occur between concurrent commitments on an account in a social network:

1. *IsFriend*: If current commitment that has arrived recently and active commitment that is running are *Friend*, both can be executed concurrent. They are consistent because they don't change the information.
2. *IsFamily*: If current commitment that has arrived recently and active commitment that is running are *Family*, current commitment would wait until active commitment execution is finished. Regards to both commitments change the information and database, if they are executed concurrently, inconsistency may happen.
3. *IsStrange*: If current commitment that has arrived recently and active commitment that is running are *Strange*, current commitment would wait until the execution of active commitment is finished. In this state if reader commitment reads the information that may be changed by writer commitment, conflict may occur.

The three states are described, are summarized in Fig.2.

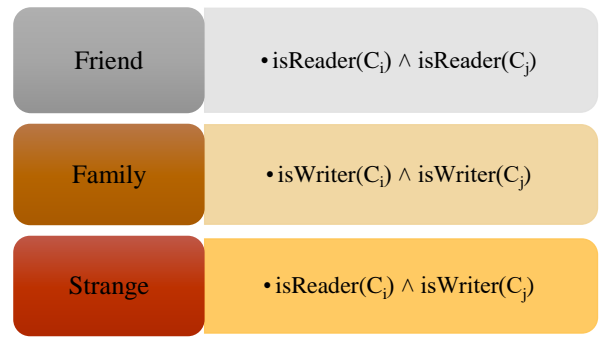

Figure 2. Commitments States

Sometimes several commitments can be created concurrently in a time slice. In this state, commitments can be executed based on two policy:

1. FCFS[1]: Commitments are serviced based on creation time in social network. This policy is fairness. In this policy commitments are pushed in waiting queue based on creation time and next commitment is selected from front of queue. So all commitments are created, can be executed.
2. Priority: Commitments are serviced by priority. The priority of commitment is assigned by social network and can change based on the social network policies. In this policy next commitment is selected based on its priority. In fact it uses a priority queue for waiting commitments.

Note that if selected commitment from waited commitments is reader, algorithm can be executed as follows:

1. In FCFS policy, all reader commitments in front of waiting queue, are selected for execution, because they are *Friend* and they can be executed concurrently.

2. In Priority policy, all reader commitments that have same priority of selected commitment, are selected for execution, because they are *Friend* and they can be executed concurrently.

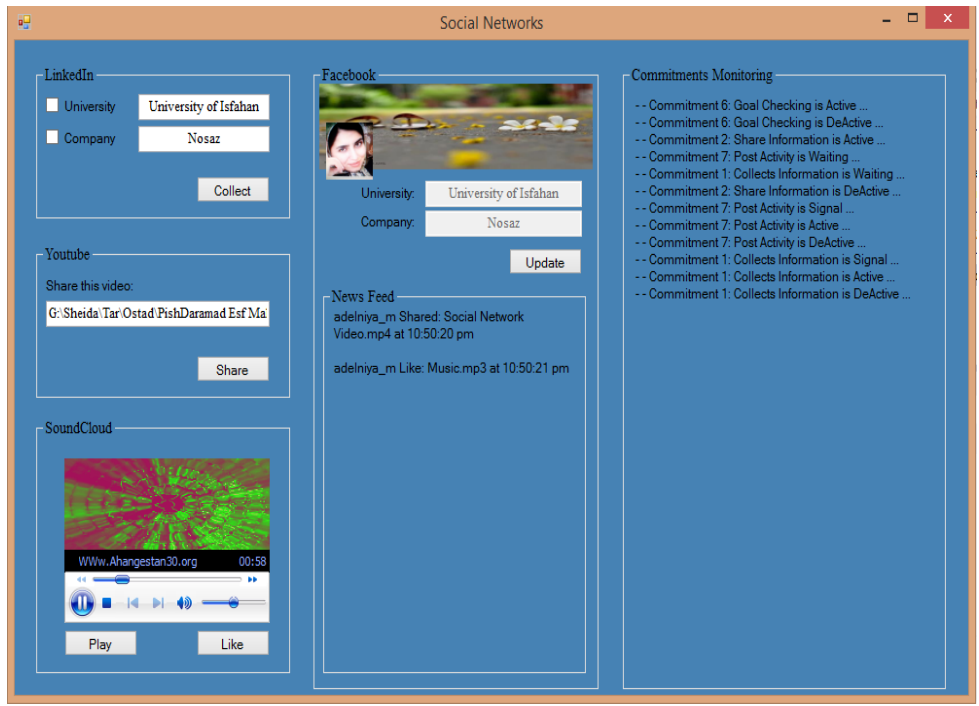

Figure 3. Implementation UI

## VII. IMPLEMENTATION

This paper uses the application designed for suggestion algorithm in previous research [3]. The user interface application is shown in Fig 3. In this application five sections have been considered as follows:

- LinkedIn: This section simulates LinkedIn social network. This social network receives some information like contact list from Facebook social network.
- YouTube: This section simulates YouTube social network operation. In this simulation, the user can share the videos on Facebook social network.

- SoundCloud: It is a music social network. Some activity like play music, like and share music on other social network can be done in it.
- Facebook: This section is central social network in this implementation that communicates with other social networks. The user activities in other social network can receive and display in this section. When social web services sign up in Facebook social network, commitments are created and consistency guarantor algorithm is executed.
- Commitments Monitoring: For monitoring the state of the commitments in any time, this section is designed and shows the number and state of all commitments for all active social web services in a social network.

---

[1] First Come, First Serve

The algorithm is checked with Facebook wall posts datasets. Experiments show that the algorithm is efficient for ensuring consistency between commitments in social web services. Also based on the experiments, the number of *Friend*, *Family* and *Strange* states are completely haphazard and depend on the time and order of incoming social web services in social networks. Figure 4 illustrates this result. Also the number of commitments that are waited in queue showed in Figure 5.

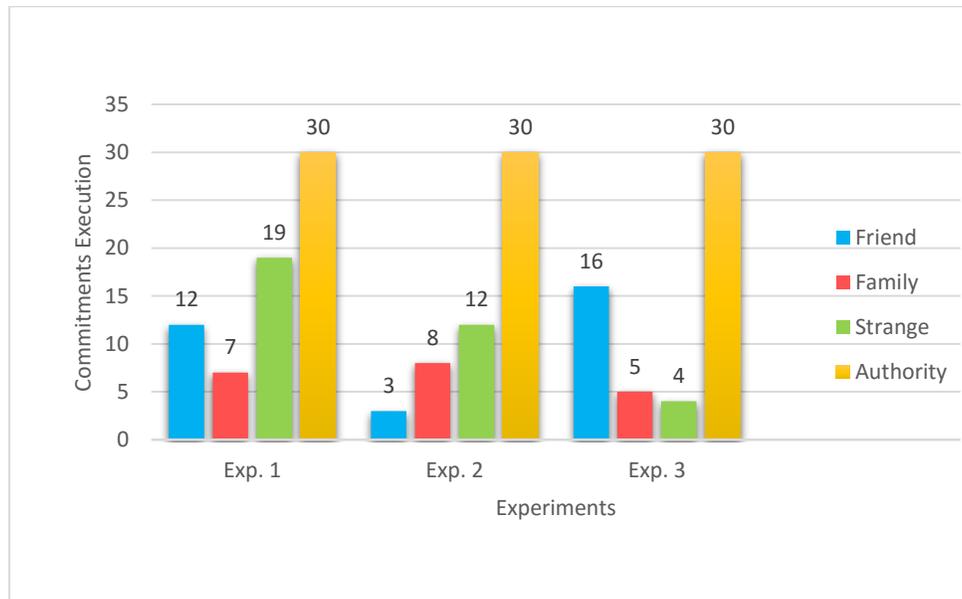

Figure 4: Number of States between Commitments in 3 Experiments

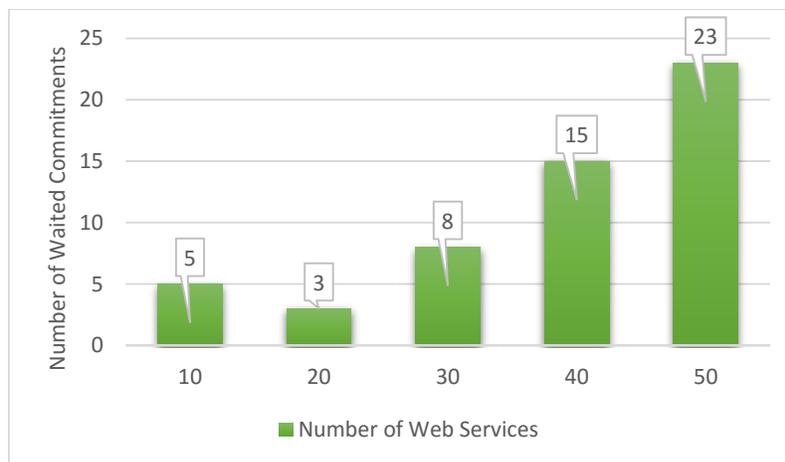

Figure 5: Number of Waited Commitments in Five Experiments

When a web service signs up in a social network, if it is authenticated by an authority component, it will be registered in the social network and changed to a social web service. Based on its operation, one or several responsibilities are assigned to this social web service. For each responsibility, commitments are defined that can accomplish tasks of social web services. For a user account in social networks, if no commitment is active, commitments of responsibility could be active and execute their operations. But if another commitment is active on this user account, consistency must be protected. Thus the algorithm checks the state of current commitments towards active commitments and decides whether commitments should be executed or waited.

For example in Fig.3, user activity are as follows:

1. Share a video on Facebook social network
2. Like a music on SoundCloud social network
3. Retrieve some information from Facebook Social network for LinkedIn account

The order of commitment execution is:

1. $C_{Resp6}$
2. $C_{Resp2}$
3. $C_{Resp7}$
4. $C_{Resp1}$

First user shares a video on Facebook social network and then like a music on SoundCloud social network and collect information from his account of Facebook. If these actions are executed in a time slice, consistency must be guaranteed.

Regards to first $C_{Resp6}$ and $C_{Resp2}$ for information sharing are created, first, goal of social web service is checked and then social web service shares the video. Also $C_{Resp7}$ and $C_{Resp1}$ have been created and are waiting for execution. Because of $C_{Resp2}$ is active commitment, $C_{Resp7}$ and $C_{Resp2}$ are *Family*. Also $C_{Resp2}$ and $C_{Resp1}$ are *Strange*. When $C_{Resp2}$ is finished, $C_{Resp7}$ is selected from waiting commitments and will be executed. After $C_{Resp7}$ execution is finished, $C_{Resp1}$ is selected from waiting queue and will be executed. Some of results showed in Table 1.

Table 1: Result of Algorithm

| Time of YouTube Sharing | Time of LinkedIn Collect Information | Time of SoundCloud Like | Time of SoundCloud Play | Status | Order of Commitments execution |
|---|---|---|---|---|---|
| Node Type | Node Type | Node Type | Node Type | | |
| 2013-06-18 16:15:07 UTC<br>Users of Youtube website | 2013-06-18 16:15:05 UTC<br>Robot | 2013-06-18 16:15:09 UTC<br>Individual Persons | | Inconsistent | $C_{Resp1}$: Collects Information is Active<br>$C_{Resp2}$: Shares Information is Waiting<br>$C_{Resp6}$: Post Activity is Waiting<br>$C_{Resp1}$: Collects Information is Deactivate<br>$C_{Resp2}$: is Signal<br>$C_{Resp2}$: Shares Information is Active<br>$C_{Resp2}$: Shares Information is Deactivate<br>$C_{Resp6}$: is Signal<br>$C_{Resp6}$: Post Activity is Active<br>$C_{Resp6}$: Post Activity is Deactivate |
| 2014-02-03 04:35:22 UTC<br>Users of Youtube website | | 2014-02-03 04:35:20 UTC<br>Individual Persons | | Inconsistent | $C_{Resp6}$: Post Activity is Active<br>$C_{Resp2}$: Shares Information is Waiting<br>$C_{Resp6}$: Post Activity is Deactivate<br>$C_{Resp2}$: is Signal<br>$C_{Resp2}$: Shares Information is Active<br>$C_{Resp2}$: Shares Information is Deactivate |
| 2014-07-18 02:21:14 UTC<br>Users of Youtube website | 2014-07-18 02:21:10 UTC<br>Robot | | 2014-07-18 02:21:18 UTC<br>SoundCloud User Account | Consistent | $C_{Resp1}$: Collects Information is Active<br>$C_{Resp1}$: Collects Information is Deactivate<br>$C_{Resp2}$: Shares Information is Active<br>$C_{Resp2}$: Shares Information is Deactivate<br>$C_{Resp6}$: Post Activity is Active<br>$C_{Resp6}$: Activity is Deactivate |

## VIII. CONCLUSION

This study set out to present an algorithm to ensure the consistency of commitments in social web services. Also two commitments are considered and added to base commitments for optimizing. In consistency guarantor algorithm, first, commitments are classified into two groups containing *Reader Commitments* and *writer commitments*. Based on this classification and inspiration of main concepts in social networks like "Family" and "Friend", the algorithm is designed. If commitments don't change the information and database, they are called *Reader Commitments* and if they affect and change information and database, they are called *Writer Commitment*. Commitments can have three states into each other. They may be *Friend*, *Family* or *Strange* based on their operations. The algorithm manages different states that may occur in commitment execution of a social web service operation. Facebook dataset is used for algorithm checking.

There are several research directions in future works of this paper. One of the future works is weaving suggested algorithm for consistency ensuring in business commitments. Also this algorithm must be checked on the professional social networks and other social networks with special performance.